\documentclass{webofc}
\usepackage[varg]{txfonts}   
\newcommand*\THP{{\sc The Three Hundred}}

\begin{document}
\title{Velocity dispersion vs cluster mass: a new scaling law
with \THP{} clusters}
%
%

\author{\firstname{Antonio} \lastname{Ferragamo}\inst{1}\fnsep\thanks{\email{ferragamoantonio@gmail.com}} \and 
	\firstname{Marco} \lastname{De Petris}\inst{1}
	\and
	\firstname{Gustavo} \lastname{Yepes}\inst{2,3}
	\and
	\firstname{Juan} \lastname{Mac\'{\i}as-P\'erez}\inst{4}
	\and
	\firstname{Weiguang} \lastname{Cui}\inst{5,3} 
	\and
	\firstname{Alejandro} \lastname{Jim\'enez-Mu\~{n}oz}\inst{4}
}

\institute{Dipartimento di Fisica, Sapienza Universit\'a di Roma, Piazzale Aldo Moro 5, I-00185 Roma, Italy  
\and
           Departamento de F\'isica T\'eorica, M\'odulo 15, 
           Facultad de Ciencias, Universidad Aut\'onoma de Madrid,
           E-28049 Madrid, Spain
\and
           Centro de Investigaci\'on Avanzada en F\'isica
           Fundamental (CIAFF), Facultad de Ciencias, Universidad Aut\'onoma de Madrid, E-28049 Madrid, Spain
\and
           Univ. Grenoble Alpes, CNRS, LPSC-IN2P3, 53, avenue des Martyrs, 38000 Grenoble, France
\and
           Institute for Astronomy, University of Edinburgh, Edinburgh EH9 3HJ, United Kingdom
          }

\abstract{%
The Planck Collaboration has shown that the number of clusters as a function of their mass and
redshift is an extremely powerful tool for cosmological analyses. However, the true cluster mass
is not directly measurable. Among the possible approaches, clusters mass could be related to different observables via self similar scaling law. These observables are related to the baryonic components of which a cluster is composed. However, the theoretical relations
that allow the use of these proxies often are affected by observational and physical biases, which impacts on the determination of the cluster mass. Fortunately, cosmological simulations are an extremely powerful tool to assess these problems. We present our calibration of the scaling relation between mass and velocity dispersion of galaxy members from the study of the simulated clusters of \THP{} project with mass above $10^{13} M_\odot$.
In order to investigate the presence of a redshift dependence, we analyzed 16 different redshifts between $z = 0$ and $z = 2$. Finally, we investigated the impact of different AGN
feedback models.
}
\maketitle
\section{Introduction}
\label{intro}
According to the hierarchical bottom-up scenario, galaxy clusters (GC) are the last structure to form, and are, therefore, the perfect tracers of the evolution of structures along the
history of the Universe. They reside in the deepest potential wells of the Large Scale Structure (LSS), generate by accretion of Dark matter (DM) \cite{springel05}, therefore their number as a function of their mass is very sensitive to the underlying Cosmology \cite{allen11}.
Although GC are dominated by DM, they include several forms and phases of the baryonic matter.
Indeed, the galactic component is surrounded by the so called Intra-Cluster Medium (ICM) composed by warm and hot gas and non-thermal plasma. Each component is characterised by its own physical process. Therefore we can measure different quantities, such as the X-ray temperature, The integrated Compton parameter or the velocity dispersion of the galaxies, by observing different regions of the electromagnetic spectrum.
Under the assumption of relaxed systems, these observables can be directly related to the cluster mass trough scaling relation. Since GC formation is gravitationaly driven starting from initial density fluctuations, they evolve self-similarly \cite{Kaiser86} and this allow us to approximate these scaling relation to power laws. This property has been shown to be fundamental for the use of clusters in cosmology, as their mass is not directly measurable. In fact, it is crucial to constrain the scaling relations as accurately and precisely as possible. For this purpose, cosmological simulations play a fundamental role being the perfect laboratories in which it is possible to calibrate and to test new scaling relations understanding possibly biases and physical limitations. 

This paper is organised as follows. In section \ref{sec:300th} we describe the simulation and the data-set we used for our analysis. In sec.~\ref{sec:scaling} we present the results of the scaling relation $\sigma-M$. Finally in section \ref{sec:conclusions} we presents our conclusions.  
\section{\THP{}}
\label{sec:300th}

The dataset of simulated GCs that we used in this study is selected within \THP{} (The300) \cite{Cui2018}.
This is a zoomed re-simulation of the 324  most massive Lagrangian region of a MultiDark Planck 2 simulation (MDPL2) \cite{Klypin2016}. The MDPL2 consists in a $1h^{-1}\mathrm{Gpc}$ volume populated with $3840^3$ DM particles of mass $1.5 \times 10^9h^{-1}\mathrm{M_\odot}$, consistent with the cosmology of Planck 2015 data release \cite{planck15_par} ($h=0.678$, $n=0.96$, $\sigma_8=0.823$,
$\Omega_\Lambda=0.693$, $\Omega_m=0.307$ and $\Omega_b=0.048$).

The region of a radius of $15h^{-1}\mathrm{Mpc}$ around the center each of the 324 selected Lagrangian regions are then populated with gas particles with initial mass of $2.36\times10^{8}h^{-1}\mathrm{M_\odot}$ and they are re-simulated with several codes. In this work we focus our analysis in the results of the Gadget-X \cite{Murante2010, rasia15} and GIZMO-SIMBA \cite{Dave2019} hydrodynamical simulations codes that implement different feedback mechanisms.

\subsection{Dataset}
\label{sec:dataset}
Within each region the bounded structures (halos and sub-halos) are identified by using the Amiga Halo Finder (AHF) \cite{AHF2009}.
For our purpose, we consider only halos with total mass $M_{200} \geq 10^{13} h^{-1}\mathrm{M_\odot}$ and with at least 5 DM sub-halos or galaxies that are not contaminated by black holes (BH) particles. The sub-halos are structure of at least 20 DM particles with mass above $10^{11} h^{-1}\mathrm{M_\odot}$. Whereas, in analogy with \cite{munari13} (hereafter M13), we call the galaxy with more than $\sim 20$ stars ($M_* \gtrsim 10^{9} h^{-1}\mathrm{M_\odot}$) within the sub-halo. 
This definition of galaxy assures the selection of all the sub-halos and many more with mass lower than $10^{11}\mathrm{M_\odot}$. 

Finally, in order to study the possible redshift evolution of the $\sigma-M$ scaling relation we selected 16 redshifts between $z=0$ and $z=2$ within the 128 snapshots stored by The300. 
 
The final selection of GC contains more than 1000 clusters at each redshift both in the Gadget-X and GIZMO catalogs.

\begin{figure*}
\centering
\includegraphics[trim=0.2cm 0.2cm 0.2cm 0.2cm, clip, width=0.49\textwidth]{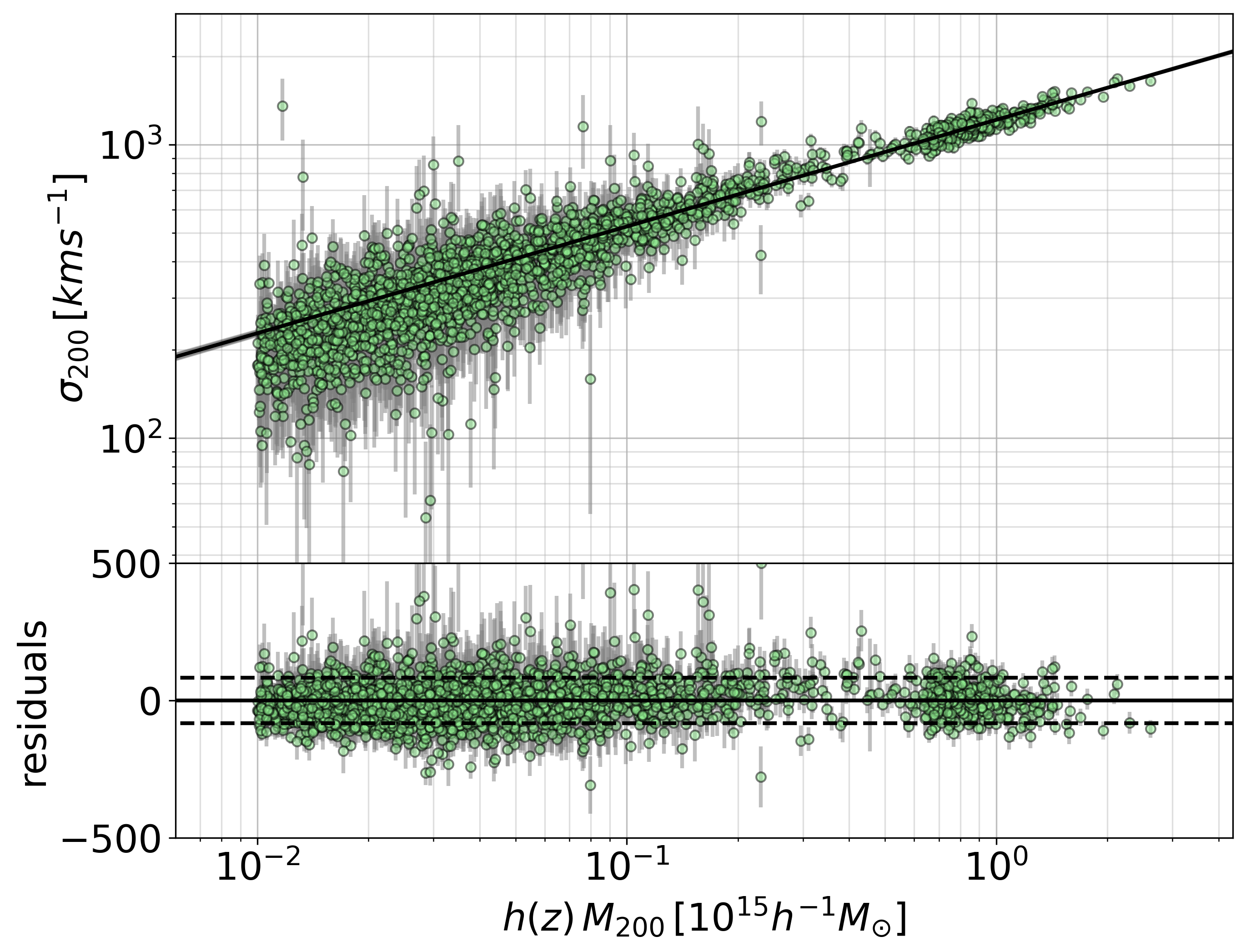}
\includegraphics[trim=0.2cm 0.2cm 0.2cm 0.2cm, clip, width=0.49\textwidth]{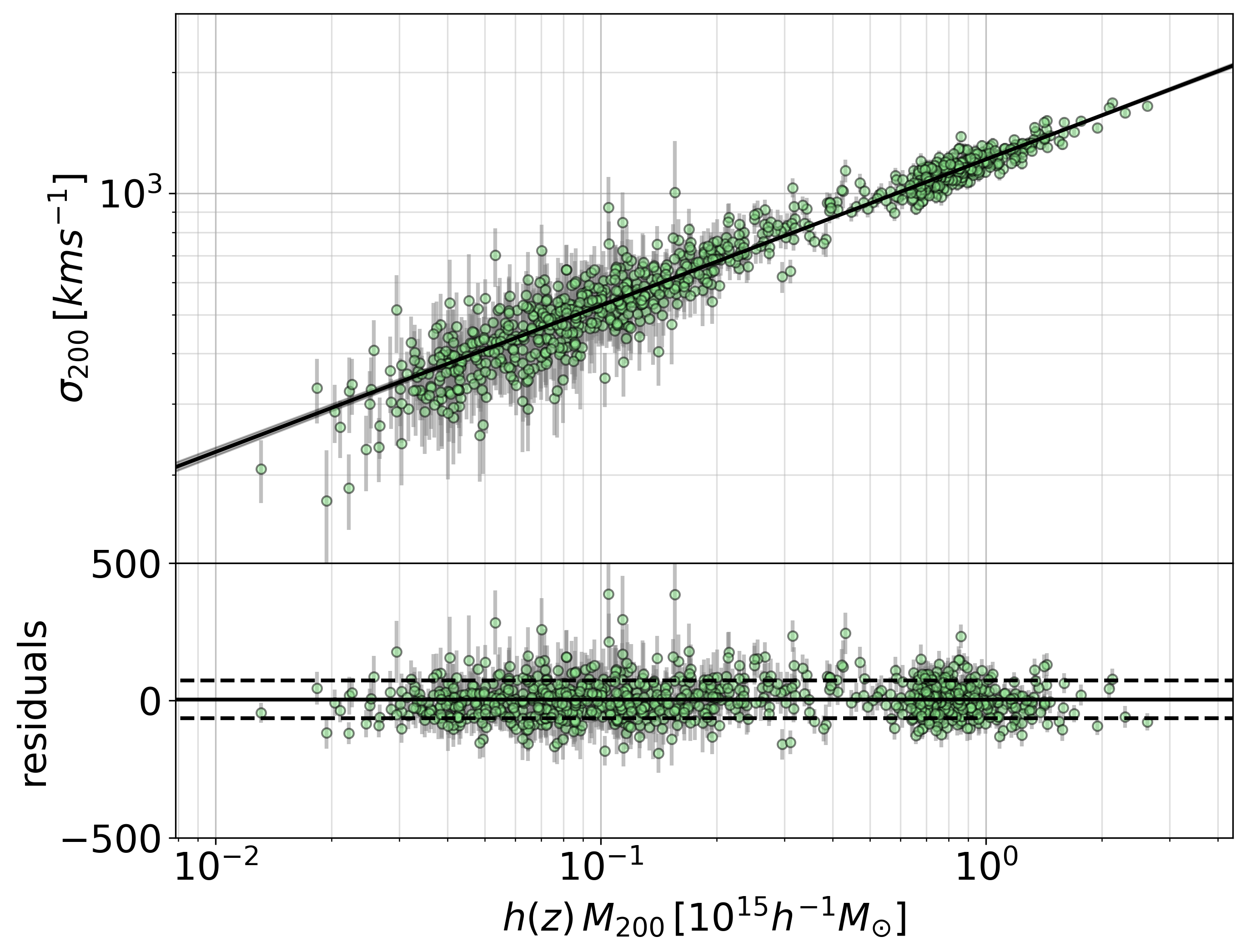}
\caption{Sub-halos velocity dispersion $\sigma_{200}$ as a function of total mass $h(z)M_{200}$, at $z=0$, for the Gadget-X simulation run. The right and left top panels show the cases $n_{gal}\geq5$ and $n_{gal}\geq10$, respectively. The solid black line represents the best-fitting relation. In the bottom panels there are depicted the corresponding residuals.}
\label{fig:sub_5_10}       
\end{figure*}

\begin{figure*}
\centering
\includegraphics[trim=0.2cm 0.2cm 0.2cm 0.2cm, clip, width=0.49\textwidth]{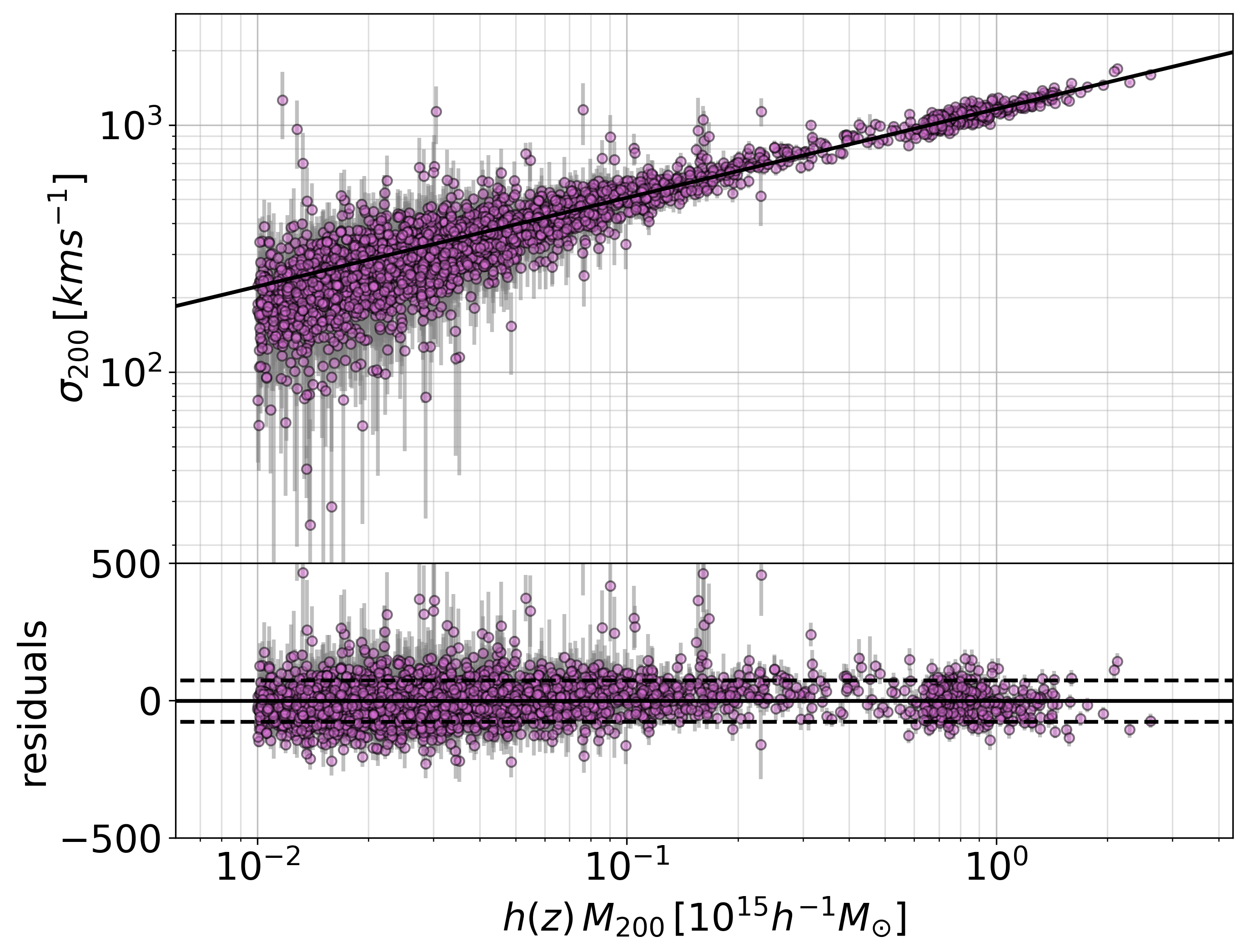}
\includegraphics[trim=0.2cm 0.2cm 0.2cm 0.2cm, clip, width=0.49\textwidth]{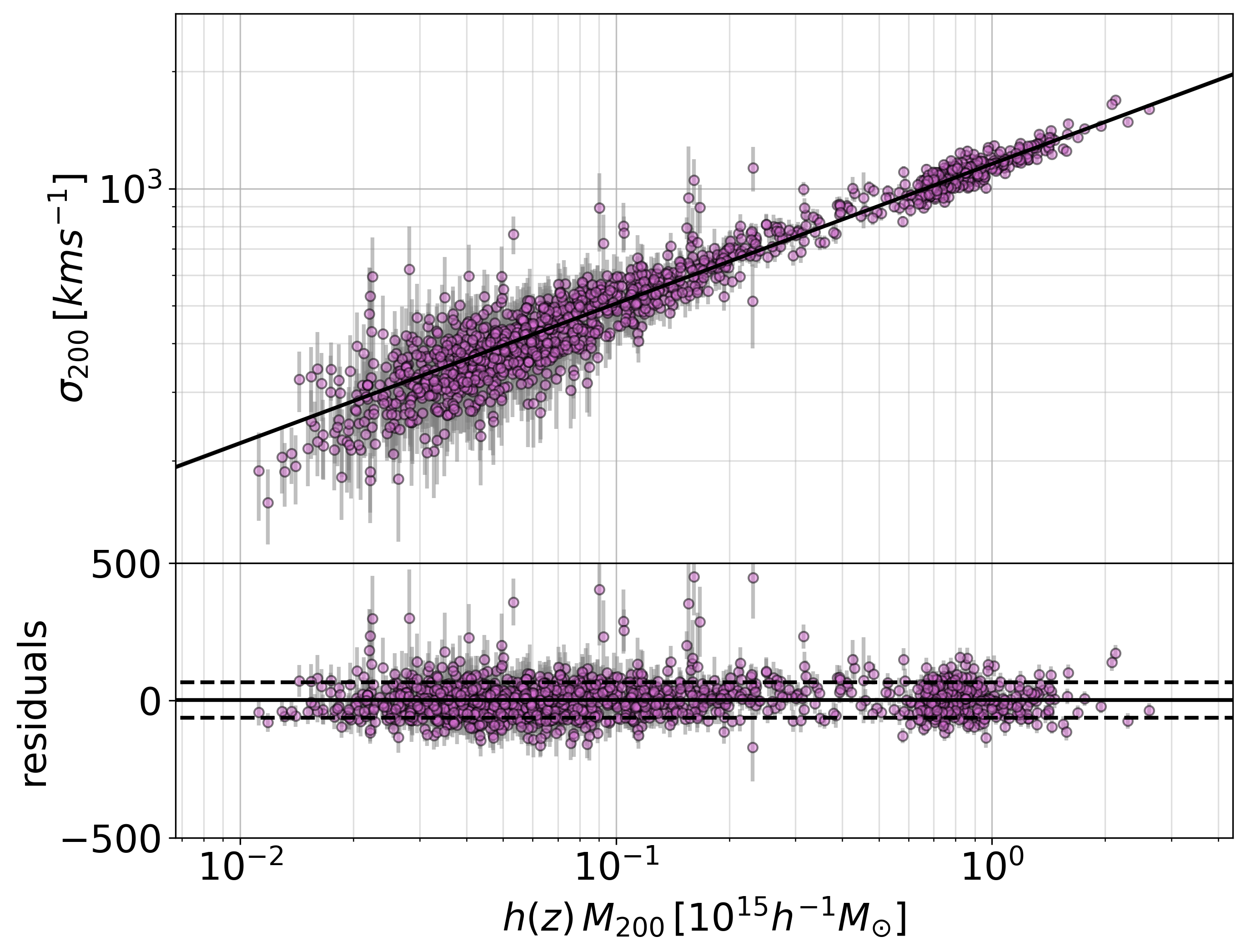}
\caption{Galaxies velocity dispersion $\sigma_{200}$ as a function of total mass $h(z)M_{200}$, at $z=0$, for the Gadget-X simulation run. The right and left top panels show the cases $n_{gal}\geq5$ and $n_{gal}\geq10$, respectively. The solid black line represents the best-fitting relation. In the bottom panels there are depicted the corresponding residuals.}
\label{fig:gal_5_10}       
\end{figure*}

\section{Scaling relation results}
\label{sec:scaling}
In analogy with the previous works in literature on the dynamical scaling relation \cite[e.g.][]{evrard08, saro13, munari13}, we characterised the parameters of a power law relation
\begin{equation}
    \frac{\sigma_{200}}{\rm km\,s^{-1}} = A\;\left(\frac{h(z)\,M_{200}}{10^{15}\,\rm M_\odot}\right)^{\alpha},
\end{equation} 
where $\sigma_{200}$ and $M_{200}$ are the velocity dispersion and the mass of the cluster within the radius $R_{200}$, respectively.
For all the clusters described in sec.~\ref{sec:dataset} we used the Biweight estimator \cite{beers90} to estimate the one-dimensional velocity dispersion $\sigma_{200}= \sigma_{200}^{3D}/\sqrt{3}$. 
Subsequently, we performed the fit with a linear function in the relation in the logarithmic space by using the python routine \texttt{scipy.optimize.curve\_fit}.

In order to investigate the impact of the number of galaxy members on the scaling relation, we performed the entire analysis taking into account two different thresholds for the minimum number of galaxy members: $n_{gal}\geq5$ and $n_{gal}\geq10$.
Figures \ref{fig:sub_5_10} and \ref{fig:gal_5_10} show the fitted relations for the Gadget-X run in the two cases at redshift $z=0$ for sub-halos and galaxies, respectively. We see that the slope of the best fit is less steep when the minimum number of galaxy members is 10. Furthermore, the scatter is higher in the case with $n_{gal}\geq5$. We can explain these evidence with the combinations of two effects. From one side the scatter increases because lower the number of galaxies higher the uncertainties on the velocity dispersion. From the other side the cluster sample in the case with $n_{gal}\geq5$ is affected by selection effects such as the Eddigton bias \cite{eddington1913}. For these reasons we present the results obtained with the sample $n_{gal}\geq10$.

\begin{figure*}
\centering
\includegraphics[trim=0.2cm 0.4cm 0.2cm 0.2cm, clip, width=0.49\textwidth]{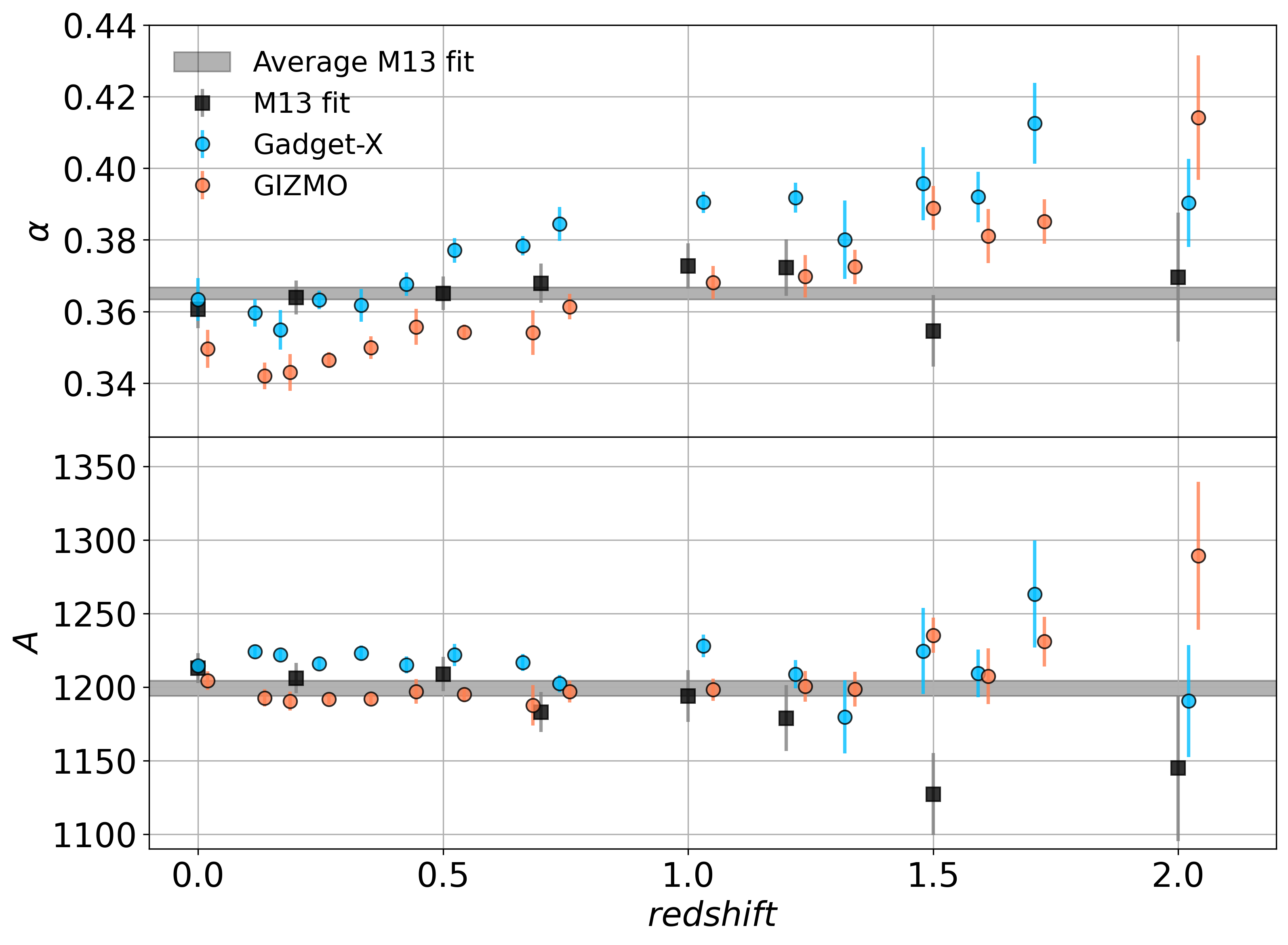}
\includegraphics[trim=0.2cm 0.4cm 0.2cm 0.2cm, clip, width=0.49\textwidth]{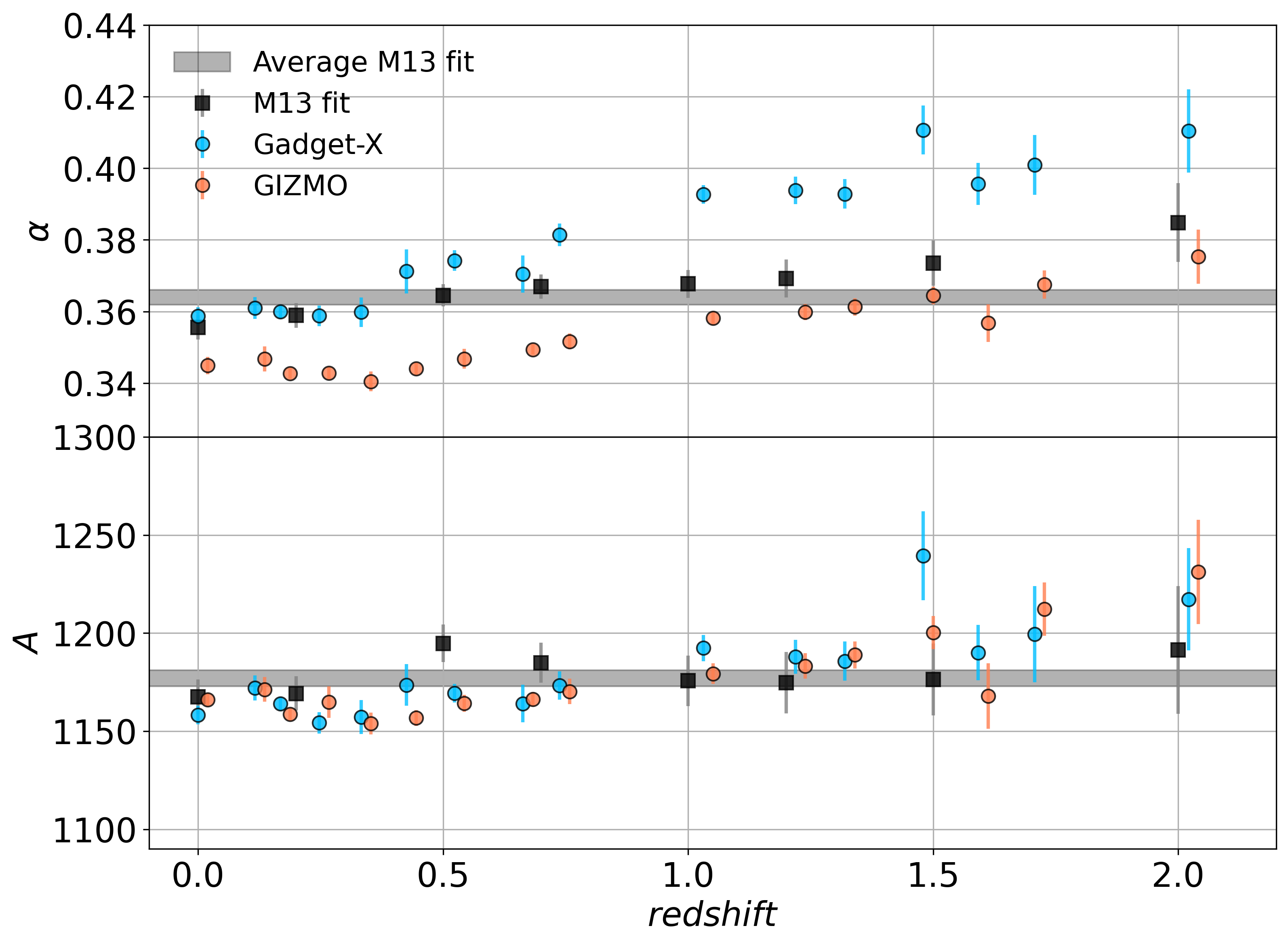}

\caption{Best fit parameters $\alpha$ (top panels) and $A$ (bottom panels) as a function of redshift, for sub-halos (left panels) and galaxies (right panels). Light blue and coral dots represent the results from the Gadget-X and Gizmo analyses, respectively. The M13 result are represented by the black squares, while the grey band represents their weighted mean.}
\label{fig:z_dep10}       
\end{figure*}

\subsection{Redshift dependence}
\label{sec:red_dep}
In figure \ref{fig:z_dep10} we show the parameter of the $\sigma-M$ relation along the redshift between $z=0$ and $z=2$ that we took into account for sub-halos and galaxies in the left and right panel, respectively. 
The black squares are the M13 results (slope in the top panel and the normalization in the central panel) for the AGN sample only at the 10 investigated redshifts, while the shaded regions represent the weighted mean of the 10 redshifts. 
In light blue and orange we plotted the best fit values obtained with the Gadget-X and the GIZMO runs of the The300. Since the Dianoga simulation and the Gadget-X implement the same baryonic physics recipe, in this section we only focus on the comparison between the M13 and Gadget-X results. The comparison with GIZMO is reported in sec.~\ref{sec:GX_G}.
In the cases of sub-halos and galaxies, the two results are in an excellent agreement at $z<0.8$. Whereas, going towards higher redshift we observe an increasing value of the slope that is not present in M13 results. 
differently from the normalization where any redshift dependence is noticeable. 

The GIZMO best fits show the same behaviour for both slope and normalization. 
For this reason we decided to take into account the redshift dependence of the slope in our parametrization of the scaling relation by fitting a linear relation between the redshift and the slope. In Tables \ref{tab:Sub-halos} and \ref{tab:Galaxies} we list the parameters of the $\sigma-M$ relation
\begin{equation}
    \frac{\sigma_{200}}{\rm km\,s^{-1}} = A\;\left(\frac{h(z)\,M_{200}}{10^{15}\,\rm M_\odot}\right)^{(\alpha+ \beta \,z)}.
    \label{eq:sc_z}
\end{equation}

\begin{table}
\centering
\caption{Best fit parameter of equation \ref{eq:sc_z} for sub-halos}
\label{tab:Sub-halos}       
\begin{tabular}{cccc}
\hline
 & A & $\alpha$ & $\beta$ \\
 \hline
Gadget-X & $1216.2 \pm 0.6$ & $0.358 \pm 0.002$ & $0.027 \pm 0.003$\\ 
GIZMO-SIMBA & $1198.8 \pm 0.7$ & $0.341 \pm 0.002$ & $0.026 \pm 0.002$ \\
\hline
\end{tabular}
\end{table}
\begin{table}
\centering
\caption{Best fit parameter of equation \ref{eq:sc_z} for galaxies}
\label{tab:Galaxies}       
\begin{tabular}{cccc}
\hline
 & A & $\alpha$ & $\beta$ \\
 \hline
Gadget-X & $1172.7 \pm 0.8$ & $0.357 \pm 0.002$ & $0.029 \pm 0.002$ \\ 
GIZMO-SIMBA & $1171.5 \pm 0.6$ & $0.340 \pm 0.001$ & $0.016 \pm 0.001$ \\
\hline
\end{tabular}
\end{table}

\subsection{Gadget-X vs GIZMO-SIMBA}
\label{sec:GX_G}
\begin{figure*}
\centering
\includegraphics[trim=0.1cm 0.1cm 0.2cm 0.2cm, clip, width=0.49\textwidth]{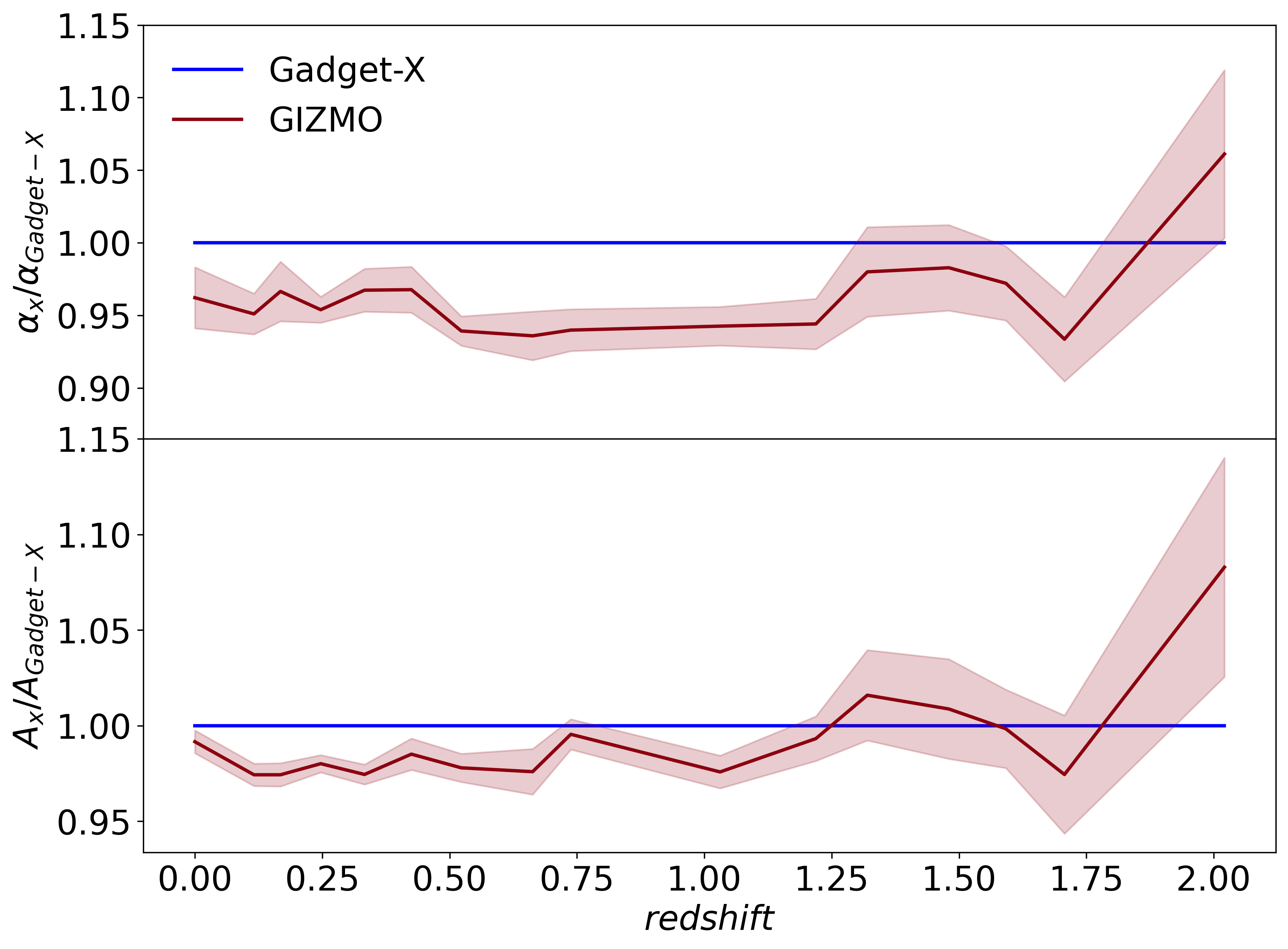}
\includegraphics[trim=0.1cm 0.1cm 0.2cm 0.2cm, clip, width=0.49\textwidth]{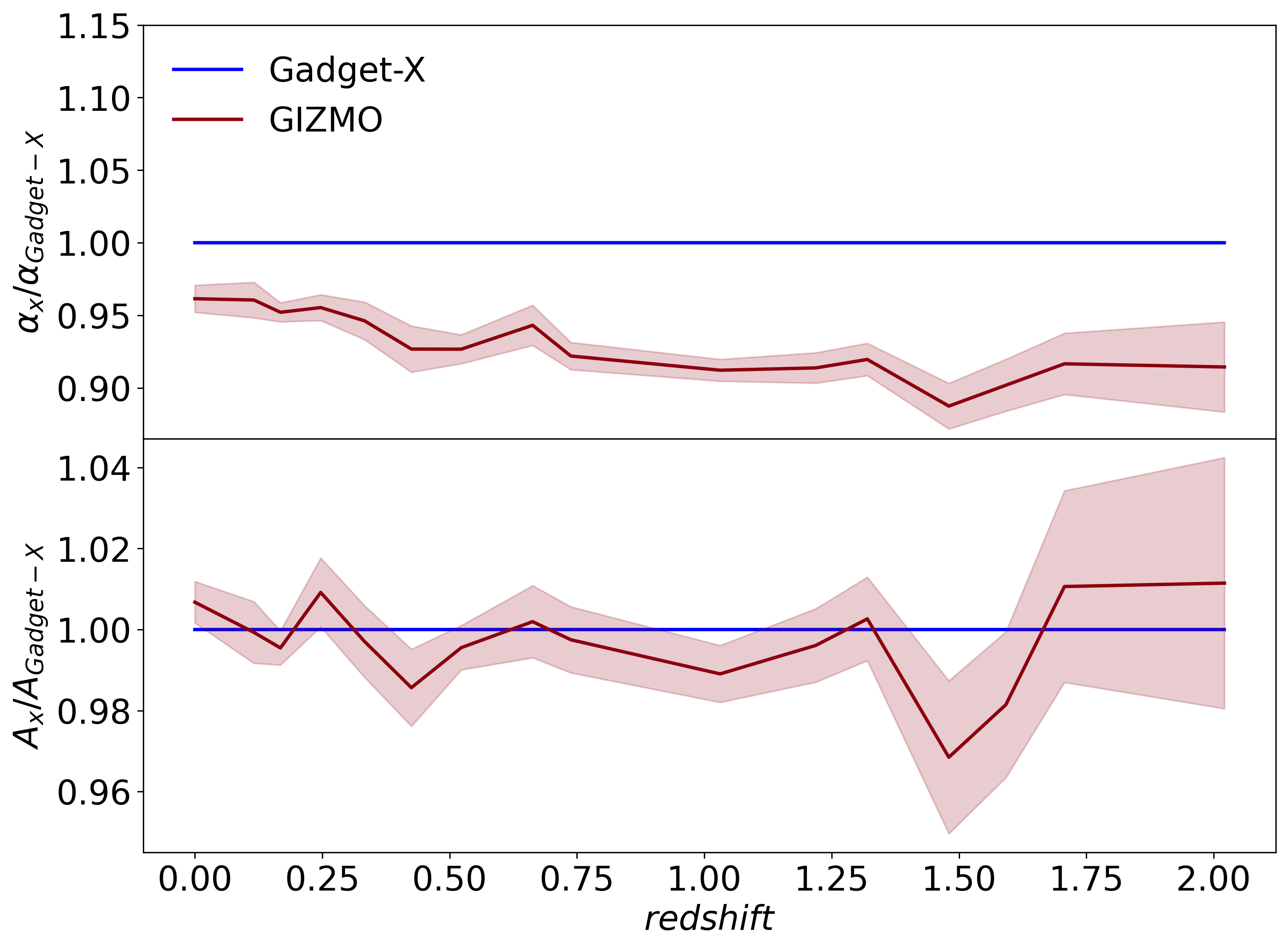}
\caption{Ratio of the best fit parameters from GIZMO and Gadget-X runs (red line), for sub-halos (left panel) and galaxies right panel). The slope, $\alpha$, and the normalization, $A$, are represented in top and bottom panels, respectively.}
\label{fig:ratio_GX_G_10}       
\end{figure*}

It is interesting to report that in Fig.~\ref{fig:z_dep10} the GIZMO parameters are always lower than the Gadget-X.
The ratio between parameters obtained from the Gadget-X and GIZMO runs are shown in Fig.~\ref{fig:ratio_GX_G_10}.
On one hand, the slope $\alpha$ from GIZMO is $\sim 5\%$ lower than the slope obtained in the Gadget-X fits in the whole redshift range for the sub-halos, whereas for the galaxies the offset increases with the redshift passing from $5\%$ at $z=0$ to a $\sim10\%$ at $z=2$. On the other hand, the bias in the normalization, $A$, is much smaller for the sub-halos and completely absent for the galaxies.
This evidence can be explained with the differences in the baryonic physics recipes between the Gadget-X and the GIZMO-SIMBA runs of The300, in particular in the differences in the AGN feedback, dramatically affecting the less massive halos, which can explain the shallower slope of GIZMO relations.

\section{Conclusions}
\label{sec:conclusions}
In this work we presented the calibration of a scaling relation between the velocity dispersion and the mass of GCs. For this purpose we used the synthetic clusters from the The300 simulation with mass $M_{200} \geq 10^{13}\mathrm{M_\odot}$, at 16 different redshifts between $z=0$ and $z=2$, from two different runs, Gadget-X and GIZMO-SIMBA.
In analogy with M13, we constrained the relation by using different tracers such as sub-halos and galaxies. As a consequence of the similarities between the Gadget-X run of The300 and the Dianoga simulation used in M13 (in particular with the AGN feedback model), we performed a direct comparison of the results of these two studies. We were able to reproduce the M13 results at redshift $z=0$. We obtained results in very good agreement with respect M13 up to $z=0.8$. After this point the two relations begin to diverge. In fact, we observed a dependence of the slope of the relation with the redshift. 
Furthermore, we investigated the effect of different model of AGN feedback on the parameters of $\sigma-M$. We compared the results from Gadget-X and GIZMO-SIMBA. The slope obtained wit GIZMO is a $\sim5\%$ smaller than the Gadget-X one for both tracers and in the whole range of redshifts.

\bibliography{biblio.bib} 

\end{document}